\documentclass[twocolumn,showpacs,preprintnumbers,amsmath,amssymb]{revtex4}
\usepackage{color}
\usepackage{graphicx}% Include figure files
\usepackage{dcolumn}% Align table columns on decimal point
\usepackage{bm}% bold math
\usepackage{epstopdf}
\begin{document}

\title{Temperature dependent graphene suspension due to thermal Casimir interaction}% Force line breaks with \\

\author{Anh D. Phan$,^1$ Lilia M. Woods$,^1$ D. Drosdoff$,^1$ I. V. Bondarev$,^2$ and N. A. Viet$^3$}

\affiliation{$^1$Department of Physics, University of South Florida, Tampa, Florida 33620, USA}
\affiliation{$^2$Physics Department, North Carolina Central, Durham, North Carolina 27707, USA}
\affiliation{$^3$Institute of Physics, 10 Daotan, Badinh, Hanoi, Vietnam}%

\date{\today} % It is always \today, today,

\begin{abstract}
Thermal effects contributing to the Casimir interaction between objects are usually small at room temperature and they are difficult to separate from quantum mechanical contributions at higher temperatures. We propose that the thermal Casimir force effect can be observed for a graphene flake suspended in a fluid between substrates at the room temperature regime. The properly chosen materials for the substrates and fluid induce a Casimir repulsion. The balance with the other forces, such as gravity and buoyancy, results in a stable temperature dependent equilibrium separation. The suspended graphene is a promising system due to its potential for observing thermal Casimir effects at room temperature.      

\end{abstract}

\pacs{Valid PACS appear here}% PACS, the Physics and Astronomy
\maketitle                                                                                                                                                                                                             

 Casimir interactions between objects arise due to electromagnetic fluctuations. The Casimir force is universal and it is present at all length scales. Investigations in the past several years directed towards understanding the fundamental nature of this interaction and its role especially in miniature devices have been particularly intense \cite{1}. In most cases, the Casimir force is attractive, and as a result close proximity between materials can lead to unwanted effects due to stiction in micro and nanomechanical systems \cite{2}. Thus, finding ways to reduce the magnitude of the force or even make it repulsive is an important field of research \cite{3,4,5,28,29,30}.    

In most vacuum separated materials, the Casimir interaction arise due to quantum mechanical effects ($T=0$ $K$). As the temperature is elevated, the force changes due to changes in the photon thermal distribution, but such effects are usually small at submicron scales and room temperature, and they are difficult to observe \cite{12,6}. 
Recently, it has been proposed to utilize objects suspended in a fluid, in which the corrections due to thermal fluctuations in the Casimir force can be observed \cite{7,8}. This method is particularly intriguing since it relies on the balance between attractive and repulsive contributions to the force arising from the dielectric response of the materials.  

Graphene is an atomically thin planar sheet of carbon atoms arranged in a honeycomb lattice. Its recent isolation has generated much scientific interest \cite{9}. The Casimir interaction involving graphene has also been considered \cite{27,10,11,16,17,24}. The unique properties of this material, originating from the linear in wave vector low energy dispersion 
 together with its 2D geometry have given rise to several unusual features. In particular, it was found that the graphene Casimir interaction is strongly temperature dependent. In the quantum mechanical limit, the force has the same distance dependence ($~d^{-4}$) as compared to the one between perfect metals, but with a significantly reduced strength \cite{4}, which is also distinct from 2D metals and insulators \cite{24}. At room temperature the interaction is dominated by thermal fluctuations in contrast to regular metals and dielectrics in vacuum, in which case it is dominated by the quantum fluctuations \cite{27,11}.    

In this Letter, we investigate a graphene sheet suspended in a fluid between substrates. We study how the Casimir interaction is influenced by the substrates. In addition, by using the balance between the Casimir force, gravity and bouyancy, we demonstrate the existence of a stable temperature dependent equilibrium of the suspended graphene.   
The chosen materials are suitable for the detection of purely thermal effects in the micron and submicron scale. This is particluarly encouraging for future experimental developments utilizing strong temperature dependent Casimir interactions.

\begin{figure}[htp]
\includegraphics[width=8.2cm]{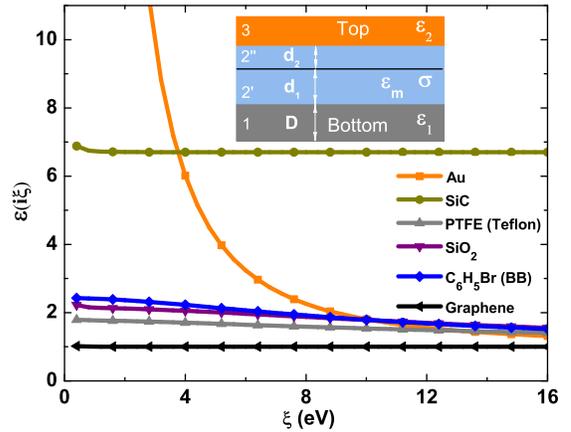}% Here is how to import EPS art
\caption{\label{fig:1}(Color online) Dielectric function $\epsilon(i\xi)$ as a function of frequency $\xi$ ($eV$) for several materials. Insert: Schematics of a planar sheet (graphene) immersed in fluid (BB) between two slab substrates (top slab - $Au$; bottom slab - $SiC$, $SiO_2$, or teflon).}
\end{figure}

The studied system is shown schematically in the insert of Fig.~\ref{fig:1}. It consists of a graphene sheet immersed in a fluid sandwiched between a lower (of thickness $D$) and an upper semi-infinite substrate. For such a layered configuration, the Casimir force per unit area can be calculated in each layer \cite{4,10,26}. The force per unit area in layer $2'$ is effectively the one between graphene and the bottom substrate given by
\begin{eqnarray}
& &F(d_{1})=-\dfrac{k_{B}T}{\pi}\sum_{n=0}^{\infty}{'} \int_{0}^{\infty}q(i\xi_{n})k_{\perp}dk_{\perp}\nonumber \\
& &\times \left(\dfrac{R_{TE}^{+}(i\xi_{n})R_{TE}^{-}(i\xi_{n})}{e^{2q(i\xi_{n})d_{1}}-R_{TE}^{+}(i\xi_{n})R_{TE}^{-}(i\xi_{n})} \right. \nonumber \\
& & \left. + \dfrac{R_{TM}^{+}(i\xi_{n})R_{TM}^{-}(i\xi_{n})}{e^{2q(i\xi_{n})d_{1}}-R_{TM}^{+}(i\xi_{n})R_{TM}^{-}(i\xi_{n})}\right),
\label{eq:1}
\end{eqnarray}
where $k_{\perp}$ is the 2D wave vector, $q (i\xi_n)= \sqrt{k_{\perp}^{2} + \varepsilon_{m}(i\xi_{n})(\xi_{n}/c)^2}$, and  $\xi_n=2n\pi k_B T/\hbar$ are the Matsubara frequencies. The prime in the sum of Eq. (1) indicates that $n=0$ term is multiplyed by 1/2.  $R_{TM}(i\xi_n)$ and $R_{TE}(i\xi_n)$ are the effective reflection coefficients for the transverse ellectric (TE) and transverse magnetic (TM) polarizations of the electromagnetic field. 

$R_{TE,TM}^{+,-}$ correspond to the effective boundary conditions due to the objects above layer $2'$ ($+$ subscipt) and below layer $2'$ ($-$ subscipt)\cite{4,26}. For the system given in Fig. 1 they are derived as follows: 
\begin{eqnarray}
& & R_{TM,TE}^{-}= r_{TM,TE}^{-}\frac{1-e^{-2k_{1}D}}{1-(r_{TM,TE}^{-})^2e^{-2k_{1}D}}, 
%R_{TE}^{-} = r_{TE}^{-}\frac{1-e^{-2k_{1}D}}{1-{r_{TE}^{-}}^{2}e^{-2k_{1}D}}, 
\nonumber\\
& & r^{-}_{TM} = \frac{{\varepsilon_{1}q - \varepsilon_{m}k_{1}}}{{\varepsilon_{1}q + \varepsilon_{m}k_{1}}},
r^{-}_{TE} = \frac{q - k_{1}}{q+ k_{1}},
\nonumber \\
& & R^{+}_{TM}= \frac{r_{TM}^{\sigma}+r_{TM}^{t}(1-2r_{TM}^{\sigma})e^{-2qd_{2}}}{1-r_{TM}^{\sigma}r_{TM}^{t}e^{-2qd_{2}}},
\nonumber \\
& & R^{+}_{TE}= \frac{r_{TE}^{\sigma}+r_{TE}^{t}(1+2r_{TE}^{\sigma})e^{-2qd_{2}}}{1-r_{TE}^{\sigma}r_{TE}^{t}e^{-2qd_{2}}},
\nonumber \\
& & r_{TM}^{\sigma} = \frac{2\pi\sigma q/\xi_{n}}{\varepsilon_{m}+2\pi\sigma q/\xi_{n}},
r_{TE}^{\sigma} = - \frac{2\pi\xi_{n}\sigma /c^2}{q+2\pi\xi_{n}\sigma /c^2}, 
\nonumber \\
& & r_{TM}^{t} = \frac{{\varepsilon_{3}q - \varepsilon_{m}k_{3}}}{{\varepsilon_{3}q + \varepsilon_{m}k_{3}}},
r_{TE}^{t} = \frac{q - k_{3}}{q + k_{3}},
\label{eq:2}
\end{eqnarray}
where $k_{1,3}=\sqrt{k_{\perp}^{2} + \varepsilon_{1,3}(i\xi_{n})(\xi_{n}/c)^2}$ and $\sigma$ is 2D graphene conductivity.

For this study, the material for the bottom substrate is  $SiO_2$, $SiC$, or teflon (PTFE), the top one is $Au$, while the medium fluid is bromobenzene (BB). We assume that the thickness of the $Au$ substrate to be much greater than the $Au$ skin depth ($\approx 22$ $nm$). Under this condition, the Casimir interaction is not influenced by the thickness, as shown by others  \cite{31}, thus we take it to be semi-infinite. $\varepsilon_{1,3,m}(i\xi)$ are taken from available experiments with Lorentz and Drude (for $Au$) models
fitted parameters \cite{3,5,13}. 

The dielectric function for graphene is found as $\varepsilon_g({\bf k},i\xi)=1+2\pi k\sigma(i\xi)/\xi$ \cite{25}. The graphene conductivity is calculated from the Kubo formalism using a two-band Dirac model \cite{14}.   
\begin{equation}
\sigma(i\xi) =\frac{2e^2k_{B}T\ln2}{\pi\hbar^2\xi} + \frac{e^{2}\xi}{\pi}\int_{0}^{\infty}\frac{\tanh[\epsilon/2k_{B}T]d\epsilon}{\epsilon^2+(\hbar\xi)^2},
\end{equation}
The first term corresponds to intraband and the second term - to interband transitions. In the infrared and low optical regime ($\leq$ 3 eV), $\sigma$ takes a universal value $\sigma_0=e^2/4\hbar$ characterizing the optical tranparency of graphene. This has been demonstrated experimentally even in room temperatures \cite{15}.  

\begin{figure}[htp]
\includegraphics[width=9.6cm]{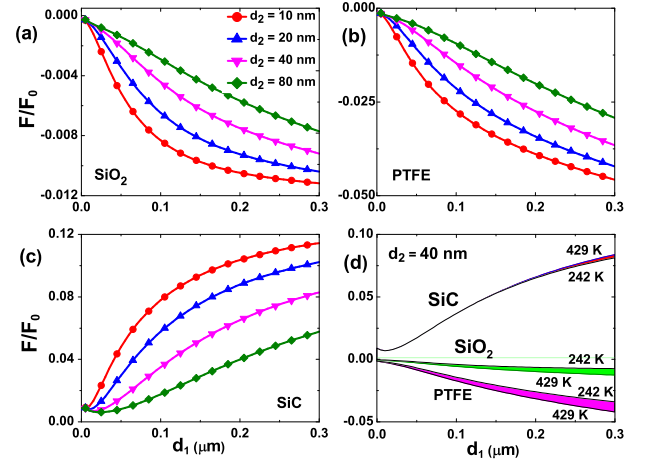}% Here is how to import EPS art
\caption{\label{fig:2}(Color online) $F/F_{0}$ as a function of  $d_1$ for (a) silica; (b) teflon; (c) $SiC$; (d) the BB liquid phase temperature regime for the three substrates. $F_{0}=\pi^2\hbar c/\left(240d_1^{4}\right)$ is the Casimir force between two perfect metal plates. The bottom substrate is semi-infinite. The BB liquid phase exists for $T=[242, 429] K$.}
\end{figure}

Using Eq.(1-3), we calculate the Casimir force between the submerged in BB graphene and the bottom substrate. The results are shown in Fig.~\ref{fig:2} (a,b,c) for several separations of the top $Au$ substrate. For $SiO_2$ and PTFE, the force is found to be repulsive, while for $SiC$ - it is attarctive for all $d_1$ and $d_2$ distances. This is understood by examining the dielectric response of the materials and their relative contribuion in the reflection coefficients. For silica and teflon and $\xi<10$ $eV$,  $\varepsilon_1(i\xi)<\varepsilon_m (i\xi) <\varepsilon_{eff}(i\xi)$, where $\varepsilon_{eff}$ is the effective dielectric function for graphene, bromobenzene in layer $2"$ and the $Au$ substrate. Since $\varepsilon_{g}$ and $\varepsilon_{BB}$ are much smaller than $\varepsilon_{Au}$, the effective materials response above layer $2'$ is  reduced as compared to the one for $Au$. This particular ordering in $\varepsilon$ \cite{5,6} results in $R_{TM}^{-}<0$, thus $F$ changes sign. Stronger repulsion is achieved by making $R_{TM}^{-}$ more negative either by bringing the $Au$ substrate closer and/or by choosing PTFE instead of $SiO_2$. For $SiC$, however, this ascending in dielectric response ordering is not maintained as $\varepsilon_{SiC}>\varepsilon_{BB}$, and one obtains attraction for all separations. We also find that for $d_2>80$ $nm$, the influence of the $Au$ is minimal and the force is changed little upon taking away the top substrate.

Graphene has an important effect on the temperature dependence of the Casimir interaction. Thermal effects in most materials become important at separations larger than the characteristic length $\lambda=\hbar c/(k_{B}T)$. For graphene, however, $\lambda$ is reduced by the fine structure constant $\alpha = 1/137$ \cite{27}, thus at room temperature thermal fluctuations effects are seen at distances greater than $25-30$ $nm$. In that case, the interaction is dominated by the $n=0$ term in Eq.(\ref{eq:1}). This is typical for graphene/graphene \cite{10} or graphene/other materials interactions \cite{4,11}. The results shown in Fig.\ref{fig:2} were performed using the full expression Eq.(\ref{eq:1}), although for $d_{2}>0.1$ $\mu m$, the $n=0$ dominates the force. In Fig.\ref{fig:2} (d), we also show how the force changes for the temperature region where BB liquid phase exists. The force modulations are not very significant, although $F$ at higher $T$ is more repulsive for PTFE and $SiO_{2}$ and more attractive for $SiC$.  

The repulsive interaction together with the strong thermal effects in the Casimir force can be used to create equilibrium separations of the submerged in BB graphene sheet above PTFE or $SiO_2$ substrates. Besides the Casimir interaction for the configuration in Fig. 1 (insert), the graphene experiences other forces - buoyancy force and gravity attraction. To illustrate the equilibrium configuration, we focus on the calculation of the total energy of the suspended graphene:
\begin{eqnarray}
& & E(d_{1}) = E_c+E_g+E_{b} \nonumber\\
& & E_{c} = \frac{k_{B}T}{2\pi}\sum_{n=0}^{\infty}{'}\int_{0}^{\infty} k_{\perp}dk_{\perp}\left(\ln\left[1-R_{TM}^{+}R_{TM}^{-}e^{-2qd_{1}}\right]\right. \nonumber \\
& & \left.+\ln\left[1-R_{TE}^{+}R_{TE}^{-}e^{-2qd_{1}}\right]\right).
\label{eq:5}
\end{eqnarray}
where $E_c$ is the Casmir energy found from Eq. (10. The energy due to gravity is $E_g=\rho_{g}g d_{1}$ with $\rho_{g} = 7.6\times10^{-7}$ $kg/m^2$ being the surface mass density of graphene\cite{18}. The buoyancy energy is $E_{b} = -\rho_{b}gN_{0}Vd_1 $, where $\rho_{b} = 1.5\times 10^{-3}$ $kg/m^{3}$ is the BB volume mass density, $g$ is the gravitational acceleration, $N_{0}$ is the number of carbon atoms per unit area and $V$ is volume of a carbon atom.

\begin{figure}[htp]
\centering
\includegraphics[width=8.2cm]{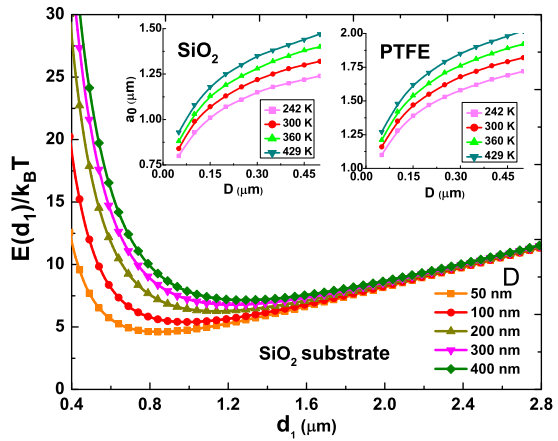}
\caption{\label{fig:4}(Color online) (Top) Total energy of the suspended graphene in terms of $k_BT$ as a function of $d_1$ for various thicknesses of the bottom substrate. Inserts show the equilibrium distance $a_0$ vs the thickness $D$ for silica and teflon for different temperatures.}
\end{figure}

We consider the case when the top substrate is not present. For distances at the submicron scale, gravity becomes important attracting the graphene sheet downward. The buoyancy acts upward together with the Casimir repulsion. The balance between these forces is captured by calculating the total energy from Eq.(\ref{eq:5}). Results for a $50\times 50$ $\mu m^2$  graphene sheet are shown in Fig.\ref{fig:4}, where the equilibrium graphene/substrate separation $a_0$ corresponds to the minimum of $E/(k_BT)$ vs $d_1$. $a_0$ is in the submicron range and it becomes larger as the thikness of the bottom substrate is increased. This behavior is also shown in the insert of Fig.\ref{fig:4} for different temperatures. One finds that for PTFE, $a_0=1.1-2.0$ $\mu m$ , while for $SiO_2$  $a_0=0.8-1.5$ $\mu m$ for 242-420 $K$. We note that the Brownian motion will cause random fluctuations around $a_0$. As a result, it is possible for stiction to occur. The probability for such a process is $\sim e^{-E/(k_BT)}$ \cite{7,8}. For the suspended graphene, the probablity is small, thus the Brownian motion is not expected to be important.  

For graphene separations in the submicron and micron scales and large temperatures, the contribution to the Casimir energy $E_C$ comes almost entirely from the $n=0$ term in the Matsubara summation due to the reduced characteristic length $\lambda$, as discussed earlier. We consider a bottom substrate with width $D=800$ $nm$. Evaluating the $n=0$ term in Eq.(\ref{eq:5}), we find $E_c=k_{B}T\gamma/(4\pi d_{1}^2)$, where $\gamma \approx 0.028$ is a constant. Note that $\gamma=0.038$ for $D=\infty$. Using this dominant term together with Eq.(\ref{eq:5}), we can describe the graphene/substrate energy around the equilibrium distance $a_0$ by a Taylor series of $E$. 

\begin{figure}[htp]
\centering
\includegraphics[width=8.2cm]{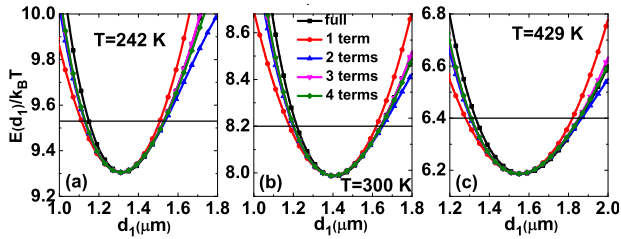}
\caption{\label{fig:5}(Color online) Total energy of the $50 \times 50$ $\mu m^2$ suspeded graphene in terms of $k_BT$ vs $d_1$ calculated by the full expression from Eq.(\ref{eq:5}) and by the first four terms in its Taylor series for $D=800$ $nm$.}
\end{figure}

In Fig.\ref{fig:5}, we show $E/(k_BT)$ as a function of the separation $d_1$ calculated via the full expression (Eq.(\ref{eq:5})) and the Taylor series expansion by retaining the first several
terms.  Fig.\ref{fig:5} shows that in a rather wide range around the equilibrium, the total energy can be described by just the first term in the Taylor series as $E(d_{1}) - E(a_{0}) \approx \frac{3k_BT\gamma}{4\pi a_0^{4}}(d_1 - a_0)^2$,
where $E(a_0)$ is the total energy at the equilibrium distance, found to be $a_{0}=\left({k_{B}T\gamma}/{2\pi(\rho_g g-\rho_bgN_0V)}\right)^{1/3}$.

This result is especially useful in relating the Casimir interaction to a simple harmonic-like oscillatory behavior of the submerged graphene. From Eq.(\ref{eq:7}), one finds that the frequency of oscillations is
\begin{equation}
\omega \sim \sqrt{\frac{3k_{B}T\gamma}{2\pi\rho_{g}a_{0}^4 }}.
\label{eq:7} 
\end{equation}
The frequency is in the typical for a mechanical system kHz regime (about 2 to 5 kHz) for the studied temperature regime and substrates.

As the graphene/substrate distance is decreased, the Casimir repulsion becomes dominant and $E$ rapidly increases. As the separation increases, the gravity attraction dominates and $E$ increases again. The balance between these three forces results in an equilibrium distance between the submerged graphene with respect to the substrate.

This study shows that graphene experiences suspension in bromobenzene above substrates providing the dielectric functions of the layers (insert Fig. 1) are in ascending (discending) order.  The suspension is characterized by a temperature-dependent equilibrium separation in the micron range. The balance between gravity, buoyancy, and the Casimir force yields one equilibrium $a_0$ which changes at a rate $~2-3$ $nm/K$ - a substantial rate at the room temperature. 

We compare this setting to the sphere/plate geometry in Ref.\cite{7, 8}, proposed for achieving tempereture-dependent Casimir effect. It was shown that a temperature dependent stable equilibrium of the suspended sphere due to the Casimir force alone is obtained by making the integrand in Eq. (1) oscillatory. This is done by choosing materials with dielectric crossings occuring at sufficiently small Matsubara frequencies close to room temperature. Note that here temperature fluctuations appear together with the quantum mechanical contributions. Inlcuding gravity and buoyancy results in the appearance of additional stable or unstable equilibriua, which maybe difficult to measure experimentally due to the possibility of sticktion. At the same time, for suspended graphene the Casimir force alone does not result in an equlibrium sepration. The inlcusion of gravity and buoyancy is necessary to ensure the desired balance. Thus we find only one stable equilibrium, which is $\sim T^{1/3}$. Another important point is that due to the significantly reduced  thermal characteristic length of graphene, only thermal fluctuations at the micron and submicron range are relavant for room temperatures. This is truly unique since in regular dielectrics and metals such effects are seen at much larger $T$ and $d_1$. We note that the suspended in fluid sphere as well as the suspended in fluid graphene attest to the richness of the temperature dependent Casimir phenomenon, which is yet to be explored experimentally.     

Furthermore, the fact that the graphene Casimir force is essentially thermal for such separations and small $T$ yields a relatively simple estimate for the oscillatory-like frequency around the stable separation. Around the equilibrium the suspended graphene behaves like a strongly temperature-dependent Hooke-like spring with characteristics directly related to the thermal Casimir force. This system is a example, which can be used by experimentalists to measure submicron separation changes over a few hundred Kelvin temperature regimes entirely due to thermal Casimir effects. Such distance scale measurements are feasible with available techniques, such as traditional setups for Casimir measurements \cite{19}, setups using suspended Michelson interferrometers \cite{20}, total internal reflection microscopy in suspensions \cite{21}, or other techniques based on transition electron microscopy, suited for oscillatory measurements \cite{22,23}.

The authors are indebted to Dr. Alexandro Rodriguez (MIT) for discussions. Financial support from the Department of Energy under Contract No. DE-FG02-06ER46297 is also acknowledged. I.V.B. was supported by the NSF-HRD-0833184 and ARO-W911NF-11-1-0189 grants. N.A.V. was supported by the Nafosted Grant No. 103.06-2011.51.

\end{document}